\documentstyle[12pt]{article}
\def\beq{\begin{equation}}
\def\eeq{\end{equation}}

\def\vcb{V_{cb}}
\def\mvcb{|V_{cb}|}
\def\be{\mbox{\boldmath $\epsilon$}}

\def\ap#1#2#3 {Ann. Phys. (NY) {\bf#1} (19#2) #3}
\def\apj#1#2#3 {Astrophys. J. {\bf#1} (19#2) #3}
\def\apjl#1#2#3 {Astrophys. J. Lett. {\bf#1} (19#2) #3}
\def\app#1#2#3 {Acta. Phys. Pol. {\bf#1} (19#2) #3}
\def\ar#1#2#3 {Ann. Rev. Nucl. Part. Sci. {\bf#1} (19#2) #3}
\def\cpc#1#2#3 {Computer Phys. Comm. {\bf#1} (19#2) #3}
\def\err#1#2#3 {{\it Erratum} {\bf#1} (19#2) #3}
\def\ib#1#2#3 {{\it ibid.} {\bf#1} (19#2) #3}
\def\jmp#1#2#3 {J. Math. Phys. {\bf#1} (19#2) #3}
\def\ijmp#1#2#3 {Int. J. Mod. Phys. {\bf#1} (19#2) #3}
\def\jetp#1#2#3 {JETP Lett. {\bf#1} (19#2) #3}
\def\jpg#1#2#3 {J. Phys. G. {\bf#1} (19#2) #3}
\def\mpl#1#2#3 {Mod. Phys. Lett. {\bf#1} (19#2) #3}
\def\nat#1#2#3 {Nature (London) {\bf#1} (19#2) #3}
\def\nc#1#2#3 {Nuovo Cim. {\bf#1} (19#2) #3}
\def\nim#1#2#3 {Nucl. Instr. Meth. {\bf#1} (19#2) #3}
\def\np#1#2#3 {Nucl. Phys. {\bf#1} (19#2) #3}
\def\pcps#1#2#3 {Proc. Cam. Phil. Soc. {\bf#1} (#2) #3}
\def\pl#1#2#3 {Phys. Lett. {\bf#1} (19#2) #3}
\def\prep#1#2#3 {Phys. Rep. {\bf#1} (19#2) #3}
\def\prev#1#2#3 {Phys. Rev. {\bf#1} (19#2) #3}
\def\prl#1#2#3 {Phys. Rev. Lett. {\bf#1} (19#2) #3}
\def\prs#1#2#3 {Proc. Roy. Soc. {\bf#1} (19#2) #3}
\def\ptp#1#2#3 {Prog. Th. Phys. {\bf#1} (19#2) #3}
\def\ps#1#2#3 {Physica Scripta {\bf#1} (19#2) #3}
\def\rmp#1#2#3 {Rev. Mod. Phys. {\bf#1} (19#2) #3}
\def\rpp#1#2#3 {Rep. Prog. Phys. {\bf#1} (19#2) #3}
\def\sjnp#1#2#3 {Sov. J. Nucl. Phys. {\bf#1} (19#2) #3}
\def\spj#1#2#3 {Sov. Phys. JEPT {\bf#1} (19#2) #3}
\def\spu#1#2#3 {Sov. Phys. Usp. {\bf#1} (19#2) #3}
\def\zp#1#2#3 {Zeit. Phys. {\bf#1} (19#2) #3}

\begin{document}
\begin{titlepage}
\begin{center}
{\Large \bf Theoretical Physics Institute \\
University of Minnesota \\}  \end{center}
\vspace{0.3in}
\begin{flushright}
TPI-MINN-97/08-T \\
UMN-TH-1534-97 \\
April 1997
\end{flushright}
\vspace{0.4in}
\begin{center}
{\Large \bf  Bound on $V+A$ admixture in the $b \to c$ current from
inclusive vs. exclusive semileptonic decays of $B$ mesons. \\}
\vspace{0.2in}
{\bf M.B. Voloshin  \\ }
Theoretical Physics Institute, University of Minnesota, Minneapolis, MN
55455 \\ and \\
Institute of Theoretical and Experimental Physics, Moscow, 117259
\\[0.2in]

\end{center}

\begin{abstract}
An admixture of a right-handed $b \to c$ current in the semileptonic
weak decays of the $B$ mesons would give a significantly different
contribution to the inclusive rate of the decays $B \to l \, \nu \, X_c$
as compared to the exclusive decay $B \to D^* \, l \, \nu$ at zero
recoil. Thus
a difference in the values of $\mvcb$ extracted from the data on these
two types of decay would measure such admixture. The present marginal
mismatch in determination of $\mvcb$ by the two methods can be
interpreted as corresponding to $g_R/g_L \approx 0.14 \pm 0.18$.
\end{abstract}
\end{titlepage}

The weak decays of $b$ hadrons are one of the most sensitive places,
where a non-standard physics may show up. The dominant weak interaction
in these decays is strongly suppressed by the small mixing parameter
$\mvcb \approx 0.04$, thus if there are new physics effects that are not
directly proportional to the weak mixing, their relative strength would
be most significant in the $b$ decays. The subject of the present note
is a bound on possible admixture of a $V+A$ current $g_R \, (\bar c_R \,
\gamma_\mu \, b_R )$ to the standard $V-A$ current $g_L \, (\bar c_L \,
\gamma_\mu \, b_L )$ in the semileptonic decays $b \to c \, l \, \nu$.
The possibility of a presence of such non-$(V-A)$ structures is most
extensively explored for the muon decay$^{\cite{pdg}}$, while for the
$b$ decays so far only the maximal case of a $(V+A) \times (V-A)$
structure of the four-fermion interaction is excluded
experimentally$^{\cite{l3,gw}}$ and another extreme of a purely vector
$b \to c$ current is clearly excluded by the very fact of non-zero
amplitude of the decay $B \to D^* \, l \, \nu$ at zero recoil.
A small value of $g_R/g_L$ is not ruled out so far and can be sought for
as one possible sign of new physics.

The structure of the four-fermion interaction for a semileptonic decay
of $b$ discussed here can be written as\footnote{It is of course also
possible to include terms where one or both leptons are right-handed.
However, since the interference of such terms with the dominant one is
suppressed by the lepton mass, only for the decay with the $\tau$ lepton
such effects may significantly show up in linear order through
interference. On the other hand the data on the semileptonic $b$ decays
with the $\tau$ are not yet sufficient for this type of a new physics
search.}
\beq
L_W=2 \sqrt{2}\, G_F \, \vcb \, \left [ (\bar c_L \, \gamma_\mu \, b_L)
+ \xi \, (\bar c_R \, \gamma_\mu b_R) \right ] \, (\bar e_L \gamma_\mu
\, \nu_L)
\label{lw}
\eeq
with $\xi = g_R/g_L$.

Since experimentally the information on polarization of the charmed
quark is not readily available, one has to seek for bounds on $\xi$ from
rate measurements. For a sensitivity to small $\xi$ it is advantageous
to look for the effects in the rate of the interference between the
right- and left- handed currents. In the decays of the $b$ hadrons the
interference of chiral components of the produced charmed quark is
proportional to the ratio of the quark mass to its energy: $m_c/E_c$.
Thus a variation of the rate between the parts of the spectrum with fast
and slow charmed quarks would give an information on the value of $\xi$.
It is the variation of the rate, rather than the total rate itself that
is informative, since an overall change in the decay rate reduces to a
shift in the extracted value of $\mvcb$. However the theoretical
predictions from QCD for the spectra are less justifiable than for the
total rates. It would be especially unreliable to extract the energy
spectrum of the charmed quarks from the data due to essential dependence
on a model for quark fragmentation into hadrons. Even for the directly
measurable charged lepton spectrum both perturbative and
non-perturbative effects are quite essential, especially near the
high-energy endpoint$^{\cite{bsuv,bksv}}$. Moreover, sufficiently below
this endpoint the effect of the $V+A$ admixture is quite uniform and
produces a suppressed effect on the variation.

It is the purpose of this note to point out that a quite sensitive to an
admixture of a $V+A$ quark current is a comparison between the inclusive
semileptonic decay rate of the $B$ mesons and the amplitude of the
exclusive decay $B \to D^* \, l \, \nu$ with zero velocity of the $D^*$.
There is a significant difference in the dependence of these two
quantities on $\xi$ in eq.(\ref{lw}), which can be used for deriving a
bound on the admixture of the right-handed current. In other words, any
mismatch between the values of $\mvcb$ extracted from the total
semileptonic decay rate and the one derived from the data on the
exclusive decay with slow $D^*$ mesons can be interpreted in terms of
the parameter $\xi$.

In order to quantify this remark, we write the expression for the total
rate of the decay $B \to l \, \nu \, X_c$ generated by the Lagrangian of
eq.(\ref{lw}) ignoring the mass of the lepton $l$\,\footnote{The data on
the decays with the $\tau$ lepton, whose mass can not be neglected, are
far less accurate than with the electron or the muon.}:
\beq
\Gamma (b \to c \, l \, \nu) = {G_F^2 \, m_b^5 \, \mvcb^2 \over 192 \,
\pi^3} \, \left [ (1+\xi^2) \, \eta_1 \, f(x)-\xi \, \eta_2 \, h(x)
\right ]~~,
\label{gl}
\eeq
where $x=m_c/m_b$, the function $f(x)$  is the standard ``blocking
factor"
\beq
f(x)=1 - 8\, x^2 + 8\, x^6 - x^8 - 24\, x^4  \ln{x}~~,
\label{f}
\eeq
and $h(x)$ describes the interference term:
\beq
h(x)=4\, x \, \left ( 1+ 9\, x^2 - 9 \, x^4 - x^6 + 12 \, x^2 \, (1+x^2)
\, \ln{x} \right )~~.
\label{h}
\eeq
Finally, the factors $\eta_1$ and $\eta_2$ describe the QCD effects. The
standard factor $\eta_1$ is studied in detail both perturbatively and
non-perturbatively (for reviews see \cite{bigi,neubert,bsu}). The factor
$\eta_2$ for the interference term is not known yet. However there is no
reason to expect any dramatic difference of $\eta_2$ from $\eta_1$. At
the present level of accuracy of other factors involved in the present
analysis it is well appropriate to approximate $\eta_2$ by $\eta_1$. In
this approximation the hypothetical admixture of the $V+A$ current in
the Lagrangian of eq.(\ref{lw}) results in the multiplicative factor in
the total semileptonic rate given by
\beq
r= \left ( 1 - {\xi \over 1+\xi^2} \, {h(x) \over f(x)} \right ) \approx
\left ( 1 - 0.74 \, {\xi \over 1+\xi^2} \right )~~,
\label{r}
\eeq
where a realistic value $x \approx 0.3$ is used in the last expression.
Thus when the mixing parameter $\vcb$ is extracted from the data on the
inclusive semileptonic decay rate, it is related to the ``true" $\vcb$
of eq.(\ref{lw}) as $\mvcb_{incl} \approx \mvcb \, (1-0.37 \, \xi)$,
assuming that $\xi$ is small.

The modification of the total inclusive semileptonic decay rate is to be
compared with modification of the amplitude of the exclusive decay
$B \to D^* \, l \, \nu$ at zero velocity $v$ of the $D^*$ in the rest
frame of the $B$ meson. This amplitude at $v=0$ is determined only by
the axial current:
\beq
\langle D^*(\be, \, v=0)|\, (\bar c \, \gamma_\mu \, \gamma_5 \, b)\, |
B \rangle = F(0) \, \epsilon_\mu ~~,
\label{me}
\eeq
where $\epsilon$ is the polarization vector of the $D^*$ and a
non-relativistic normalization for the states is used. Thus in
accordance with the coefficient of the axial current in eq.(\ref{lw})
the amplitude gets the factor $(1-\xi)$ in the presence of a $V+A$
admixture. The form factor $F(0)$ is theoretically
tractable$^{\cite{vs}}$ within the heavy quark theory. In the leading
approximation $F(0)=1$ and both the perturbative corrections and the
non-perturbative ones due to finite masses of the $c$ and $b$ quarks are
being extensively studied over last years (see the recent review papers
\cite{bigi,neubert,bsu}).
The current ``best value" for $F(0)$ with all corrections included is
\beq
F(0)=~~0.90 \pm 0.03^{\cite{bigi}},~~0.91 \pm 0.03^{\cite{neubert}},~~
0.91 \pm 0.06^{\cite{bsu}}~~.
\label{f0}
\eeq
Clearly, there is a consensus on the central value, and the difference
in the estimate of the theoretical uncertainty depends on how much
conservative attitude is chosen. Most likely the value 0.03 is a fair
estimate of the uncertainty, while 0.06 should be regarded as its
conservative maximum.

An extrapolation of the experimentally measured amplitude to the point
of zero recoil of the $D^*$ yields the value of $\mvcb_{excl} \,
|F(0)|$, where $\mvcb_{excl}$ is thus extracted ``exclusive" value of
the weak mixing parameter. In the presence of the $V+A$ admixture in the
current this quantity is related to the ``true" $\mvcb$ as $\mvcb_{excl}
= \mvcb \, |1-\xi|$. Thus there emerges a difference between
$\mvcb_{excl}$ and $\mvcb_{incl}$, given for small $\xi$ by
\beq
{\mvcb_{incl} \over \mvcb_{excl}} \approx 1+0.63 \, \xi~~.
\label{rv}
\eeq

The current status of the experimental data and the theoretical
determination of $\mvcb$ from the inclusive and the exclusive processes
can be derived from the review papers \cite{bigi}, \cite{neubert} and
\cite{bsu}. Their results for $\mvcb_{incl} \times 10^3$ are 
respectively $41.3 \pm 1.6_{exp} \pm 2_{th}$, $40 \pm 1_{exp} \pm
4_{th}$, and $41.9 \pm 1.6_{exp} \pm 2_{th}$\footnote{The estimate of
the errors in the latter number is obtained by adding linearly the
uncertainties from all sources presented separately in Ref. \cite{bsu}
and ascribing the largest of the rest two quoted experimental errors.}.
The existing experimental data on the exclusive decay allow for a fairly
accurate extrapolation to the kinematical point of zero recoil of the
$D^*$, resulting in the current ``world average" value $\mvcb_{excl} \,
|F(0)| = (34.1 \pm 1.4) \times 10^{-3}$ (for a discussion and references
to specific experiments see Ref. \cite{neubert}). This datum and the
quoted above theoretical results for $F(0)$ correspond to the extracted
values of $\mvcb_{excl} \times 10^3$ respectively $37.7 \pm 1.6_{exp}
\pm 2_{th}$\,\footnote{This is a literal quote from Ref. \cite{bigi},
although the theoretical uncertainty in it is somewhat larger, than one
would obtain by using the discussed there uncertainty in $F(0)$.},
$37.5 \pm 1.5_{exp} \pm 1.2_{th}$, and $37.5 \pm 1.5_{exp} \pm
2.5_{th}$.

It is not entirely clear how to weigh and compare these results in order
to estimate the limits on the difference $\Delta \vcb =
\mvcb_{incl}-\mvcb_{excl}$, given possible correlations in the
experimental data on the rates of the inclusive and the exclusive
processes and also given certain correlations in the theoretical
analyses. Quite loosely, one may estimate the current mismatch between
the values of $\mvcb$ extracted by the two different methods as
$\Delta \vcb \approx (3.5 \pm 4.5) \times 10^{-3}$, with the error being
dominated by the theoretical uncertainty, and therefore not considered
as one standard deviation. This estimate translates into $\Delta \vcb/
\mvcb \approx 0.09 \pm 0.12$ and according to eq.(\ref{rv}) into an
estimate of the $V+A$ admixture parameter $\xi \approx 0.14 \pm 0.18$.
Clearly, for a more accurate estimate a dedicated analysis of the
experimental data and of the theoretical uncertainties specifically
aimed at determining $\Delta \vcb$ is appropriate.

It is interesting to notice, that a value $\xi \approx 0.14$ would
reduce, according to eq.(\ref{r}) the semileptonic branching ratio for
the $B$ mesons by about 10\% of its value. This reduction may be quite
welcome in view of a possible problem of a low $B(B \to l \, \nu \, X)\,
^{\cite{bbsv}}$.

It would be far beyond the scope of the present note to discuss possible
new physics sources of the $V+A$ admixture in eq.(\ref{lw}). Still, it
can be noticed that in view of the recent experimental indication from
the experiments at HERA$^{\cite{h1,zeus}}$ of a leptoquark with mass
about 200 GeV and a coupling to $e \bar u$ or to $e \bar d$, $g \sim
0.03 - 0.05\, ^{\cite{bcjw}}$, one may speculate that an exchange of a
scalar coupled to both $(l \bar b)$ and $(\nu \bar c)$ with similar mass
and coupling would result in an effective admixture of the $V+A$
structure in eq.(\ref{lw}) with $\xi \sim 0.1$. Needless to mention
however the complications$^{\cite{bcjw}}$ with the flavor changing
neutral currents that arise from such scalars.

In summary. It is pointed out that the difference in the values of
$\mvcb$ extracted from the total inclusive semileptonic decay rate of
the $B$ mesons and from the amplitude of the exclusive decay $B \to D^*
\, l \, \nu$ is sensitive to an admixture of a right-handed $b \to c$
current in the effective Lagrangian for semileptonic $b$ decays. The
results of the existing analyses of $\mvcb_{incl}$ and $\mvcb_{excl}$
can be interpreted in terms of the admixture parameter as corresponding
to $\xi \approx 0.14 \pm 0.18$. A dedicated analysis of the bounds on a
mismatch between these two values of $\mvcb$ can further improve bounds
on the presence of the $V+A$ structure.

I am thankful to Mikhail Shifman, Nikolai Uraltsev and Yuichi Kobota for
useful discussions of theoretical and experimental aspects of the
problem discussed here. This work is supported, in part, by the DOE
grant DE-AC02-83ER40105.


\begin{thebibliography}{99}
\bibitem{pdg}
Particle Data Group, R.M. Barnett {\it et.al.}, \prev{D54}{96}{1}.
\bibitem{l3}
L3 Coll. M. Acciarri {\it et.al.}, \pl{B351}{95}{375}.
\bibitem{gw}
M. Gronau and S. Wakaizumi, \prl{68}{92}{1814}.
\bibitem{bsuv}
I. Bigi, M. Shifman, N. Uraltsev, and A. Vainshtein, \prl{71}{93}{496}.
\bibitem{bksv}
B. Block, L. Koyrakh, M. Shifman, and A.I. Vainshtein,
\prev{D49}{94}{3356}.
\bibitem{bigi}
I. I. Bigi, Report UND-HEP-96-BIG06, November 1996; [hep-ph/9612293].
\bibitem{neubert}
M. Neubert, Report CERN-TH/97-24, February 1997; [hep-ph/9702375].
\bibitem{bsu}
I. Bigi, M. Shifman, and N. Uraltsev, Report TPI-MINN-97/02-T,
UMN-TH-1528-97, UND-HEP-97-BIG01, March 1997; [hep-ph/9703290].
\bibitem{vs}
M.B. Voloshin and M.A. Shifman, Sov.J.Nucl.Phys. {\bf 47}, 511 (1988).
\bibitem{bbsv}
I. Bigi, B. Block, M.A. Shifman, and A.I. Vainshtein,
\pl{B323}{94}{408}.
\bibitem{h1}
H1 Coll.  C. Adloff {\it et.al.}, Report DESY-97-024, February 1997;
[hep-ex/9702012].
\bibitem{zeus}
ZEUS Coll. J. Breitweg {\it et.al.}, Report DESY-97-025, February 1997;
[hep-ex/9702015].
\bibitem{bcjw}
K. S. Babu, C. Kolda, J. March-Russell, and F. Wilczek, Report
IASSNS-HEP-97-04, March 1997; [hep-ph/9703299].
\end{thebibliography}
\end{document}